\pdfoutput=1
\documentclass[%
 reprint,
superscriptaddress,
a4paper,
longbibliography,
amsmath,amssymb,
numbers,
prb,
ulem
]{revtex4-1}

\usepackage[pdftex]{hyperref,graphicx}
\usepackage[usenames, svgnames]{xcolor}
\usepackage{amsfonts,amssymb,amsmath,mathrsfs}
\usepackage[separate-uncertainty=true]{siunitx}
\usepackage[english]{babel}
\usepackage[utf8]{inputenc}
\usepackage[T1]{fontenc}
\usepackage{xspace}
\usepackage{upgreek}
\usepackage[normalem]{ulem}
\usepackage{comment}
\usepackage{xfrac}
\def\ket#1{\mathinner{|{#1}\rangle}}

\usepackage{braket}

\newcommand{\figref}[2]{\hyperref[#1]{\getrefnumber{#1}(#2)}}

\hypersetup{
 pdftitle = {Observation of topological transport quantization by dissipation in fast Thouless pumps},
 pdfauthor = {Zlata Fedorova, Haixin Qiu, Stefan Linden, Johann Kroha},
	colorlinks=true,linkcolor=blue,citecolor=blue,filecolor=blue,urlcolor=blue,pdfpagemode=UseNone,pdfstartview={XYZ null null 1.00}
}

\addto\captionsenglish{}
\addto\captionsenglish{}

\renewcommand\textemdash{\leavevmode\unskip\kern0.8pt\rule[0.19\baselineskip]{8pt}{0.4pt}\kern1pt\ignorespaces}

\begin{document}

\title{Observation of topological transport quantization by dissipation in fast Thouless pumps}

\author{Zlata Fedorova}
\email{cherpakova@physik.uni-bonn.de}
\affiliation{Physikalisches Institut, Rheinische Friedrich-Wilhelms-Universit\"at Bonn, Nussallee 12, 53115 Bonn, Germany.}
\author{Haixin Qiu}
\email{haixin@th.physik.uni-bonn.de}
\affiliation{Physikalisches Institut, Rheinische Friedrich-Wilhelms-Universit\"at Bonn, Nussallee 12, 53115 Bonn, Germany.}
\author{Stefan Linden}
\email{linden@physik.uni-bonn.de}
\affiliation{Physikalisches Institut, Rheinische Friedrich-Wilhelms-Universit\"at Bonn, Nussallee 12, 53115 Bonn, Germany.}
\author{Johann Kroha} 
\email{kroha@th.physik.uni-bonn.de}
\affiliation{Physikalisches Institut, Rheinische Friedrich-Wilhelms-Universit\"at Bonn, Nussallee 12, 53115 Bonn, Germany.}

\date{\today}

\begin{abstract}
Quantized dynamics is essential for natural processes and technological applications alike. The work of Thouless on quantized particle transport in slowly varying potentials (Thouless pumping) has played a key role in understanding that such quantization may be caused not only by discrete eigenvalues of a quantum system, but also by invariants associated with the nontrivial topology of the Hamiltonian parameter space. Since its discovery, quantized Thouless pumping has been believed to be restricted to the limit of slow driving, a fundamental obstacle for experimental applications. Here, we introduce non-Hermitian Floquet engineering as a new concept to overcome this problem. We predict that a topological band structure and associated quantized transport can be restored at driving frequencies as large as the system’s band gap. The underlying mechanism is suppression of non-adiabatic transitions by tailored, time-periodic dissipation. We confirm the theoretical predictions by experiments on topological transport quantization in plasmonic waveguide arrays.
\end{abstract}

\maketitle

The standard realization for Thouless pumping\cite{Thouless1983,Niu1984} is the time-periodic version of the Rice-Mele (RM) model~\cite{Rice1982}, which describes a dimerized tight-binding chain whose system parameters change cyclically along a closed loop in Hamiltonian parameter space. 
In the adiabatic regime and for a completely filled band, the net particle transfer per cycle is an integer given alone by the Berry phase associated with the loop or the Chern number of the band, i.e., a topological invariant robust against topology-preserving deformations of the parametric loop.
Such nontrivial topology of the Hamiltonian parameter space or band structure
was recognized as the overarching concept behind phenomena apparently
as diverse as the integer quantum Hall effect~\cite{Thouless1982}, the
quantum spin Hall effect \cite{Kane2005quantumspinHall},
topological insulators in solid state~\cite{Franz-Molenkamp2013topoinsulators} and photonics~\cite{rechtsman2013photonic,hafezi2013imaging},
quantum spin~\cite{Fu2006} or charge pumping~\cite{Thouless1983},
Dirac or Weyl semimetals~\cite{Armitage2018},
and the electric polarization of crystalline solids \cite{king1993theory}.
Recently, topological or Thouless pumping was experimentally demonstrated using ultracold atoms in dynamically controlled optical lattices~\cite{lohse2016thouless,nakajima2016topological} or using waveguide arrays~\cite{kraus2012topological}. 
 
In realistic systems, however, Thouless pumping generically faces two difficulties.
First, at nonzero driving frequencies, unavoidable in experiments, the system becomes topologically trivial. The reason is that the nonzero driving frequency
defines a Floquet-Bloch Brillouin zone (FBBZ) and the dimension of
the band structure is increased by one compared to the adiabatic case. 
The coupling between forward- and backward-propagating states then opens a
gap~\cite{titum2016anomalous,lindner2017universal,privitera2018nonadiabatic}, so that the Chern number, or winding number around the FBBZ,
of the effectively two-dimensional band becomes trivial,
and the particle transport deviates from perfect quantization.
Second, realistic experimental systems are to some extent open and subject
to dissipation, so that the quantum mechanical time evolution of single-particle states deviates from unitarity, which may prevent the closing of the cycle
in Hamiltonian parameter space. This motivates the interest in
non-Hermitian (NH) Hamiltonians.
Non-Hermiticity can have profound influence on the system dynamics.
In addition to ubiquitous exponential decay, it may cause such peculiar phenomena as dissipation-induced localization in the Caldeira-Legget model \cite{caldeira1983path}, unidirectional robust transport~\cite{longhi2015robust}, asymmetric transmission or reflection~\cite{lin2011unidirectional,feng2013experimental}, or NH topological edge states associated with exceptional points~\cite{lee2016anomalous,leykam2017edge,malzard2015topologically}. Non-Hermiticity has been utilized to probe topological quantities ~\cite{rudner2009topological,ozawa2014anomalous}. Another fascinating example is the so-called non-Hermitian shortcut to adiabaticity~\cite{bender2007faster,ibanez2011shortcuts,torosov2013non}, which describes faster evolution of a wavefunction in an NH system than in its Hermitian counterpart. 

Here, we introduce time-periodic modulation of dissipation as a
new concept to restore topological transport quantization in fast Thouless
pumps. 
Although in many-body systems dissipation would be induced by interactions 
or particle loss, the plasmon polariton dynamics in our experiments is mathematically identical to that of a 
linear, dissipative, periodically driven Schr\"odinger equation.   
To analyze systems of this kind theoretically, we utilize 
the Floquet theory for non-Hermitian, time-periodic systems. 
Using this formalism, we demonstrate for a driven RM model
that time-periodic dissipation can give rise to a band structure in
the two-dimensional FBBZ with a nontrivial Chern number.
Hence, the mean displacement of a wave
packet per cycle is quantized even when the driving frequency is fast, i.e.,
far from adiabaticity. 
In a real-space picture, this topologically quantized transport comes
about, because the time-periodic loss selectively suppresses the hybridization
of a right(left)-moving mode with the counterpropagating one.
The theoretical predictions are confirmed
by experiments on arrays of coupled dielectric-loaded surface plasmon-polariton
waveguides (DLSPPW)~\cite{bleckmann2017spectral}.
DLSPPWs are uniquely suited model systems for realizing 
topological transport with dissipation: The propagation of
surface plasmon polaritons mathematically realizes the single-particle
Schr\"odinger equation on a one-dimensional tight-binding lattice
\cite{christodoulides2003discretizing,bleckmann2017spectral},
where the waveguide axis resembles time, and the
system parameters, including losses, can easily be modulated along the
waveguide axis. Moreover, complete band filling 
is achieved via Fourier transform to k-space by pumping a single site 
(waveguide) of the tight-binding lattice. This is essential for probing the
band topology which otherwise is possible only in fermionic systems
at low temperature. 

\section*{Results}
\subsection*{Model}
\begin{figure}[t!]
\includegraphics[width=\linewidth]{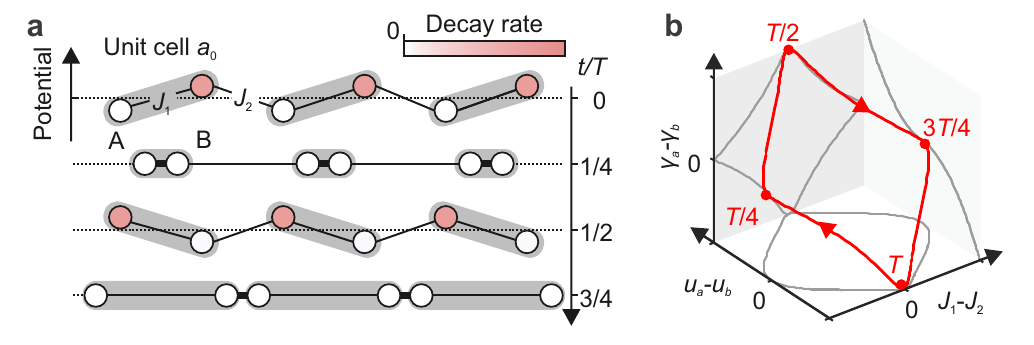}
\caption{\textbf{Non-Hermitian driven Rice-Mele model} (a)~Schematic of the periodically driven, NH RM lattice for
  four equidistant times during a pumping cycle. Lossy sites are depicted by
  red color, large (small) hopping amplitudes $J_{1,2}$ by short (long)
  distances between sites.
  (b) Pumping cycle in the parameter space $(J_1-J_2,u_{a}-u_{b},\gamma_{a}-\gamma_{b})$.
}
 \label{fig:ModelSample}
\end{figure}

We consider a periodically driven RM model ~\cite{asboth2016short,privitera2018nonadiabatic} with additional onsite, periodic dissipation (see Fig.~\ref{fig:ModelSample}),
$\hat{H}(t)=\hat{H}_{\mathrm{RM}}(t)- \mathrm{i}\hat{\Gamma}(t)$,  
\begin{eqnarray}
  \hat{H}_{\mathrm{RM}}(t)&=&{\sum}_{j}\left(J_1(t)\hat{b}^{\dagger}_j\hat{a}_j+J_2(t)\hat{a}^{\dagger}_{j+1}\hat{b}_j +h.c.\right) \label{eq:RiceMeleHamiltonian}\\
       &+&{\sum}_{j}\left( u_a(t)\hat{a}^{\dagger}_{j}\hat{a}_j+ u_b(t)\hat{b}^{\dagger}_{j}\hat{b}_j\right), \label{eq:RiceMeleHamiltonian2} \nonumber \\
\hat{\Gamma}(t)&=&{\sum}_{j}\left(\gamma_a(t)\hat{a}^{\dagger}_{j}\hat{a}_j+\gamma_b(t)\hat{b}^{\dagger}_{j}\hat{b}_j\right).\label{eq:PeriodicLoss}
\end{eqnarray}
where $j$ runs over all unit cells, $\hat{H}_{\mathrm{RM}}(t)$ is the Hamiltonian of the periodically
driven, nondissipative RM model and $\hat{\Gamma}(t)$ describes the losses.  
$\hat{a}^{\dagger}_j$ and $\hat{b}^{\dagger}_j$ ($\hat{a}^{\phantom{\dagger}}_j$
and $\hat{b}^{\phantom{\dagger}}_j$) are creation (annihilation) operators
in unit cell $j$ on sublattice $A$ and $B$, respectively.
The inter-/intra-cell hopping amplitudes, $J_{1/2}(t)$ and the onsite
potentials on the two sublattices, $u_{a}(t)$ and $u_{b}(t)$,
are all real-valued, periodic functions of time with frequency
$\Omega=2\uppi/T$ according to
\begin{align*}
&u_a(t)=-u_0 \cos(\Omega t+\varphi),  &  u_b(t)=u_a(t-T/2),\\
&J_1(t)=J_0e^{-\lambda(1-\sin \Omega t )}, &  J_2(t)=J_1(t-T/2),
\end{align*}
with $u_0,\, J_0,\, \lambda>0$, and $\varphi=0$ (unless otherwise specified).
The choice of the hopping amplitudes is motivated by 
the exponential dependence of the wave-function overlaps on the spacing
$\lambda(1-\sin{\Omega t})$ between neighboring sites, as in our experiment
below. In our NH modification of the RM model, the time-periodic decay
rates $\gamma_a(t) \geq 0$ and $\gamma_b(t) \geq 0$
are nonzero once the onsite potential exceeds the mean
value ${[u_a(t)+u_b(t)]}/{2}=0$. This resembles, for instance,
a realistic situation where particles in a trapping potential are lost
from the trap once the trapping potential is not sufficiently deep.
Thus, we choose
\begin{align*}
&\gamma_a(t)=-\gamma_0\, \Theta(u_a(t))\,\cos(\Omega t+\varphi), &  \gamma_b(t)=\gamma_a(t-T/2).
\end{align*}

\subsection*{Non-Hermitian Floquet analysis}
In the following calculations we use the non-Hermitian Floquet formalism discribed in the Methods section below.
We assume $u_0=J_0=1$, $\lambda=1.75$ and all energies are given in units of $J_0$. 
\begin{figure}[t!]
	\includegraphics[width=\linewidth]{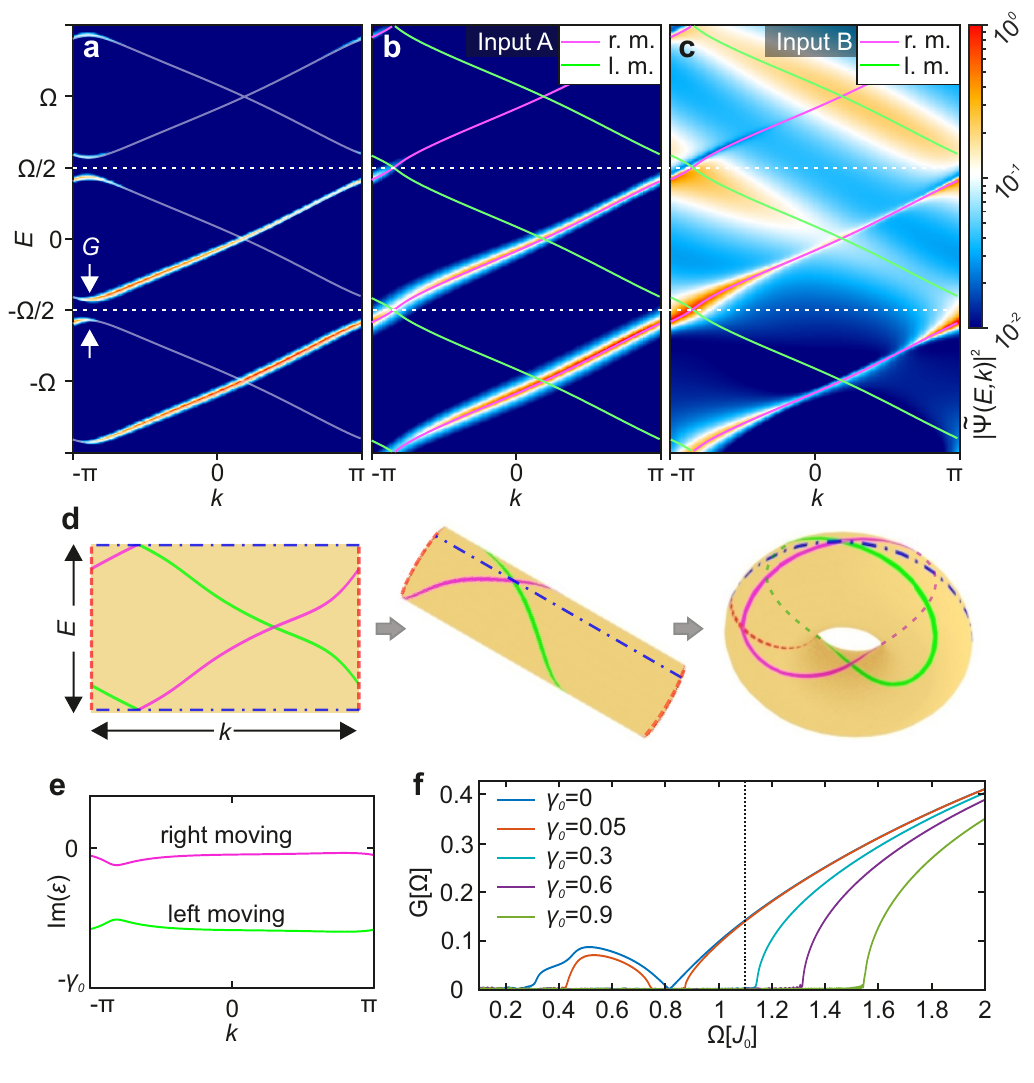}
	\caption{\textbf{Floquet analysis of the driven NH RM model in the non-adiabatic regime.} Calculated band structures of the RM model for driving frequency $\Omega=1.1J_0$. Thin lines: left- and right-moving Floquet
		quasienergy bands (real parts). (a) Band structure of the Hermitian RM
		model ($\gamma_0=0$). The band gaps at $E=\pm \Omega/2$ indicate
		a topologically trivial band structure, i.e., the breakdown of transport	quantization. 
		color code: normalized spectral occupation density of a state $\ket{\Psi(t)}$ injected at time $t=0$ on a single site of the sublattice $A$, calculated from Equation~(\ref{eq:Intensity}). It is seen that this injection almost homogeneously populates the right-moving bands, and almost no mixing of different Floquet modes occurs, as described by Equation~(\ref{eq:Intensity1}). (b) Same as in (a) but for the NH RM model with $\gamma_0=0.4J_0$ when the system is excited at a single site of the initially nonlossy sublattice $A$. As in (a), almost no mixing of Floquet modes occurs. 
		The gap at the FBBZ boundary is closed, restoring transport quantization.
		(c) Same as in (b), but for a state injected at a site of the initially
		lossy sublattice $B$. Although the band gaps remain closed by the
		dissipation, this predominantly populates
		the left-moving band with a broad distribution, and the losses are high.
		(d) The first FBBZ which evolves into a 2D torus due to the periodicity along the $E$ axis (coincidence of dashed-dotted lines) as well as the $k$ axis (coinciding dashed lines). The magenta and green lines are the forward- and backward-propagating dispersions analogous to (b) and (c). They wind around the torus with winding numbers $Z=\pm 1$ (c.f. Methods).
		(e) Imaginary part of the quasienergy bands presented in (b, c), showing low dissipatiion in the right-moving band. (f) The size of the band gap $G$ in dependence on the driving frequency at different loss amplitudes $\gamma_0$. The black dashed line shows $\Omega=1.1 J_0$.}
	\label{fig:Quasienergies}
\end{figure} 

\begin{figure}[t!]
	\includegraphics[width=\linewidth]{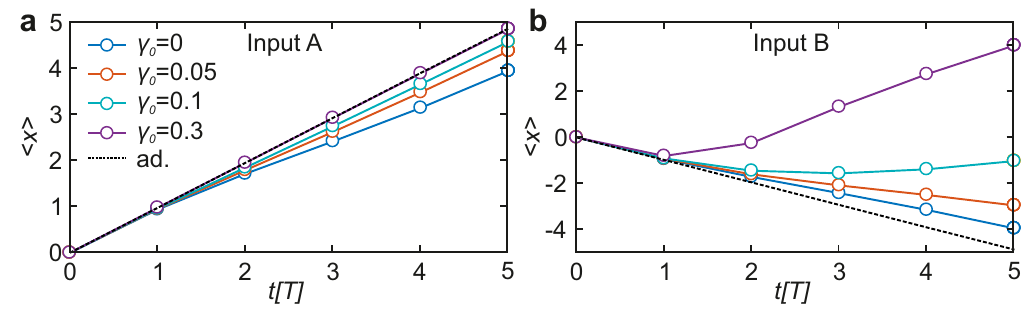}
\caption{\textbf{The center-of-mass shift in the NH RM model.} The center-of-mass position of the injected wavepacket after up to 5 full pumping cycles ($\Omega=1.1J_0$) at different loss aplitudes $\gamma_0$ for a single-site input on (a) sublattice $A$ and (b) sublattice $B$.}
\label{fig:Centerofmass}
\end{figure}

In view of the experimental setup discussed below, we also consider the
time evolution of states $\ket{\Psi_A(t)}$ and $\ket{\Psi_B(t)}$ which
have been initialized (``injected'') at time $t=0$ with nonzero amplitude
only at a single site of the $A$ or $B$ sublattice, respectively. For the parametric cycle configuration shown in Fig.~\ref{fig:ModelSample}~(b) the chosen initial time moment leads to an asymmetric amplitude distribution of the counter-propagating Floquet states with respect to the two sublattices.
As seen in Fig.~\ref{fig:Quasienergies}, such initial conditions populate, by Fourier expansion, almost homogeneously an entire
right- or left-moving band. Thus, it is a way to create the topologically
important complete band filling, which would otherwise be possible only in
fermionic systems. 
In the Hermitian case ($\gamma_{0}=0$), we see from 
Fig.~\ref{fig:Quasienergies} (a) that
the counterpropagating bands hybridize, accompanied by avoided crossings 
and gaps with width $G$ opening at the Floquet Brillouin zone boundaries, so that the
bands become topologically trivial. As a result, the charge pumped per 
period deviates from the quantized value.
This marks the generic breakdown of quantized Thouless pumping at any finite 
pumping frequncy $\Omega$, as also noted 
in \cite{titum2016anomalous,privitera2018nonadiabatic}.  Note that computing the gap size $G$, as visible in Fig.~\ref{fig:Quasienergies} (a), involves diagonalization of the entire Floquet Hamiltonian matrix. In leading order perturbation theory, $G$ would be given by the Fourier amplitude of the periodic drive, i.e., for the first FBBZ by $J_0$, which strongly differs from the exact value.

We now consider the NH RM model driven with $\gamma_0=0.4J_0$ (see Figs.~\ref{fig:Quasienergies}~(b-e)). 
Adding losses leads to several profound effects. First, the quasienergies become complex, whereby the right- and left-moving bands acquire considerably different dampings shown in Fig.~\ref{fig:Quasienergies}~(e) and seen as different
broadenings of the spectral band occupation in Fig.~\ref{fig:Quasienergies}~(b), (c). Second, the two inputs are no longer equivalent in respect to the relative populations of the two bands. In particular, for the input $A$ we almost exclusively excite right-moving states, while for the input $B$ in addition to the lossy left-moving states, we  partially populate right moving-states. Third, and most importantly, the gap $G$ closes and, hence, the bands wind around the entire 2D FBBZ as illustrated in Fig.~\ref{fig:Quasienergies}~(d). 
In the Methods section it is shown that this restores the quantized transport (see Equation~\eqref{eq:disspumping2}). 
Note, that these effects only occur once $\gamma_0$ is larger than some threshold value. In order to study this threshold behaviour we numerically evaluated the gap size $G$ at various driving frequencies and loss amplitudes (see Fig.~\ref{fig:Quasienergies}~(f)). In the Hermitian case ($\gamma_0=0$) the gap size has a complex oscillatory behaviour~\cite{avron1999quantum} as a function of the driving frequency. Our analysis shows that a larger gap size requires stronger damping in order to close it. For instance, at the previously analyzed driving frequency $\Omega=1.1J_0$ the loss amplitude $\gamma_0$ should be larger than $0.3J_0$ to close the gap.

Next, we investigate the position of the center of mass (CoM) of the wave-packet, 
$\langle x \rangle (t)=\bra{\Psi(t)}x\ket{\Psi(t)}/\braket{\Psi(t)|\Psi(t)}$, after up to 5 completed driving cycles at various losses and fixed driving frequency $\Omega=1.1J_0$ for different initial conditions input $A$ or $B$ (see Fig.~\ref{fig:Centerofmass}~(a,b)). In the adiabatic case the mean displacement is almost $+1$ ($-1$) unit cell per cycle for delta-like excitations on sublattice $A$ ($B$). Small deviations from unity result from slight inhomogeneity of the band population. At the driving frequency $\Omega=1.1J_0$ the displacement per cycle is considerably smaller in the Hermitian case ($\gamma=0$) indicating deviation from the quantized transport. With increasing losses this deviation becomes smaller and smaller for input $A$ and for $\gamma\ge 0.3$ the displacement can not be distinguished from the adiabatic case. Surprisingly, for the input $B$ we observe that the CoM position switches direction with time.
This is a signature of the chirality of the Floquet bands and is due to the fact that the propagation of even poorly populated low-loss states in positive $x$-direction starts to dominate after the first few periods, while the states propagating in negative $x$-direction are quickly damped due to the phase relation of the periodic losses with respect to the hopping amplitude.

\subsection*{Experiments}\label{sec:Experiments}

\begin{figure}[t]
\includegraphics[width=\linewidth]{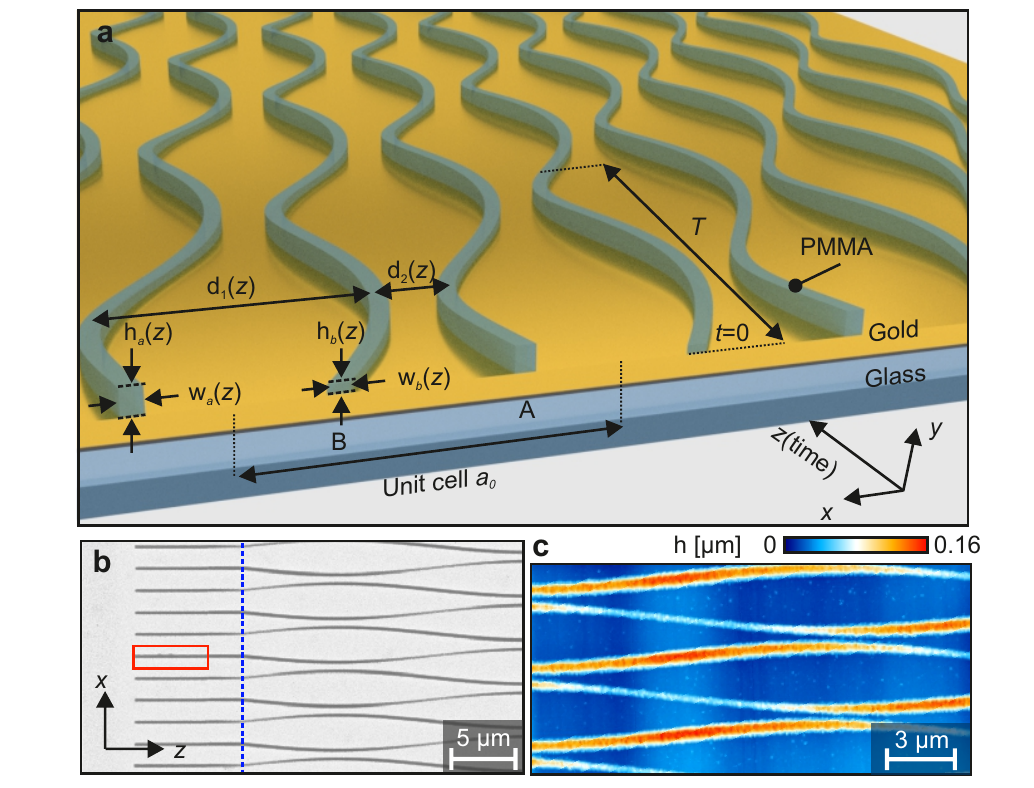}
 \caption{{\textbf{Plasmonic implementation of the NH RM model.} (a) Sketch of the plasmonic implementation of the NH RM model. (b) Scanning electron micrograph of a typical sample corresponding to $J_0=0.144\,\upmu \mathrm{m}^{-1}$, $\Omega=1.45J_0$, $u_0=1.1\,J_0$, $\gamma_0=0.8\,J_0$. The red dotted box highlights the grating coupler deposited onto the input waveguide A.  (c) AFM scan of the same sample as shown in (b).}}
 \label{fig:ArraySketch}
\end{figure}

In order to test our theoretical predictions we performed experiments based on DLSPPWs.
The experimental realization of the model described by Equations~\eqref{eq:RiceMeleHamiltonian}, \eqref{eq:PeriodicLoss} is based on the mathematical equivalence between the time-dependent Schr\"odinger equation in tight-binding approximation and the paraxial Helmholtz equation which describes propagation of light in coupled waveguide arrays~\cite{christodoulides2003discretizing,bleckmann2017spectral}.  
Figure~\ref{fig:ArraySketch} shows a scheme of a DLSPPW array (a) as well as a 
scanning electron micrograph (b) and an AFM scan (c)  of a typical sample. The sample fabrication process and the typical geometrical parameters of the arrays are described in the Methods section.
The waveguide array represents a dimerized 1D lattice, where each unit cell contains two waveguides, $A$ and $B$. Here, the propagation direction $z$ plays the role of time.
Periodic modulation of the effective hopping amplitudes is reached by sinusoidally varying the spacing between the adjacent waveguides $\mathrm{d}_{1,2}(z)$ while the on-site potential variation is realized by changing the waveguides' cross-sections (heights $\mathrm{h}_{a,b}(z)$ and widths $\mathrm{w}_{a,b}(z)$).
In addition, the variation of the waveguides' cross-section affects the instantaneous losses $\gamma_{a,b}(z)$. 
When the cross-section decreases, the confinement of the guided mode weakens. As a result, the modes can couple to free-propagating surface plasmon polaritons (SPPs) and scatter out from the array. We employ this effect to introduce time-dependent losses $\gamma_{a,b}(z)$.

\begin{figure}[hbt]
\includegraphics[width=\linewidth]{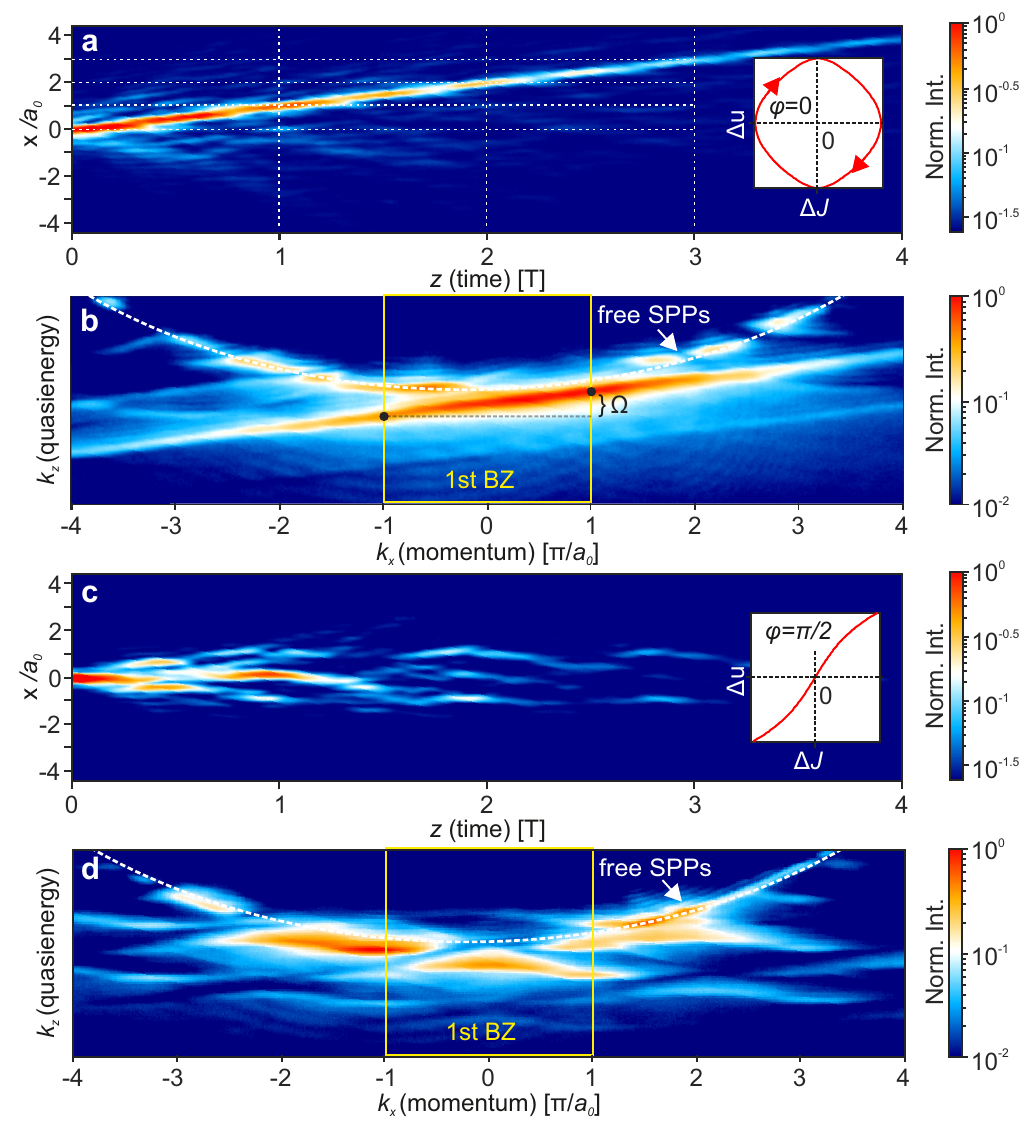}
 \caption{{\textbf{Observation of fast Thouless pumping in a DLSPPW array.} (a) Real-space SPP intensity distribution for $u_0=1.1J_0$, $\gamma_0=0.8J_0$, and $\varphi=0$. Plot on the right shows the projection of the corresponding pumping cycle onto the plane  $(J_1-J_2,u_{a}-u_{b})$. (b) Fourier-space SPP intensity distribution corresponding to (a). (c)~Real-space SPP intensity distribution for $u_0=1.1J_0$, $\gamma_0=0.8J_0$, and $\varphi=\uppi/2$. Plot on the right again shows the corresponding cycle in parameter space. (d) Fourier-space SPP intensity distribution corresponding to (c).}}
 \label{fig:Experiment_LRM}
\end{figure} 
We first consider a pumping cycle that encloses the critical point.  
For this purpose we choose the geometrical parameters of the DLSPPW array such
that $u_0=1.1J_0$ and $\Omega=1.45J_0$. By comparing the real-space intensity
distribution to numerical calculations we estimate the loss amplitude to be
$\gamma_0=0.8 J_0$. The real-space SPP intensity distribution $I(x,z)$ recorded by leakage radiation microscopy (see Methods) for single site excitation at site A is shown in Fig.~\ref{fig:Experiment_LRM}~(a). 
According to the aforementioned quantum optical analogy this corresponds to the probability density $I(x,t)=\vert \Psi(x,t)\vert^2$.
We observe for all $z$ a strongly localized wave packet, whose CoM is  transported in positive $x$-direction in a quantized manner, i.e., by one unit cell per driving period (see dotted lines), even though the driving frequency $\Omega$ is larger than the modulation amplitudes of all relevant parameters. 

The corresponding momentum resolved spectrum $I(k_x,k_z)$  is obtained by Fourier-space leakage radiation microscopy~\cite{fedorova2019limits} and is shown in  Fig.~\ref{fig:Experiment_LRM}~(b). This intensity distribution is analogous to the spectral energy density presented in Fig.~\ref{fig:Quasienergies}.
We note that this technique provides the full decomposition in momentum components in the higher Brillouin zones~\cite{bleckmann2017spectral}.
The main feature of the spectrum is a continuous band with average slope $a_0/T$.
The absence of gaps in the band indicates that the band winds around the 2D FBBZ $\{-\Omega/2\leq k_z< \Omega/2;-\uppi/a_0\leq k_x< \uppi/a_0\}$.
This is a hallmark of a quantized pumping and confirms our theoretical predictions (see Fig.\,\ref{fig:Quasienergies} (b)).

As a reference measurement, we consider the parametric cycle, where all parameters are changing with the same amplitudes as in the previous case but the phase is chosen as $\varphi=\uppi/2$. Under these conditions the Hamiltonian is symmetric under space and time inversion.
In Fig.~\ref{fig:Experiment_LRM}~(c) we present the real-space SPP intensity distribution for this parametric cycle. 
In contrast to the previous case the wave packet is spreading and we do not observe CoM transport in $x$-direction.
The corresponding momentum resolved spectrum shows a complicated band structure with multiple band gaps (see Fig.~\ref{fig:Experiment_LRM}~(d)). Obviously, none of the bands winds around the 2D FBBZ.

\begin{figure}[t]
	\includegraphics[width=\linewidth]{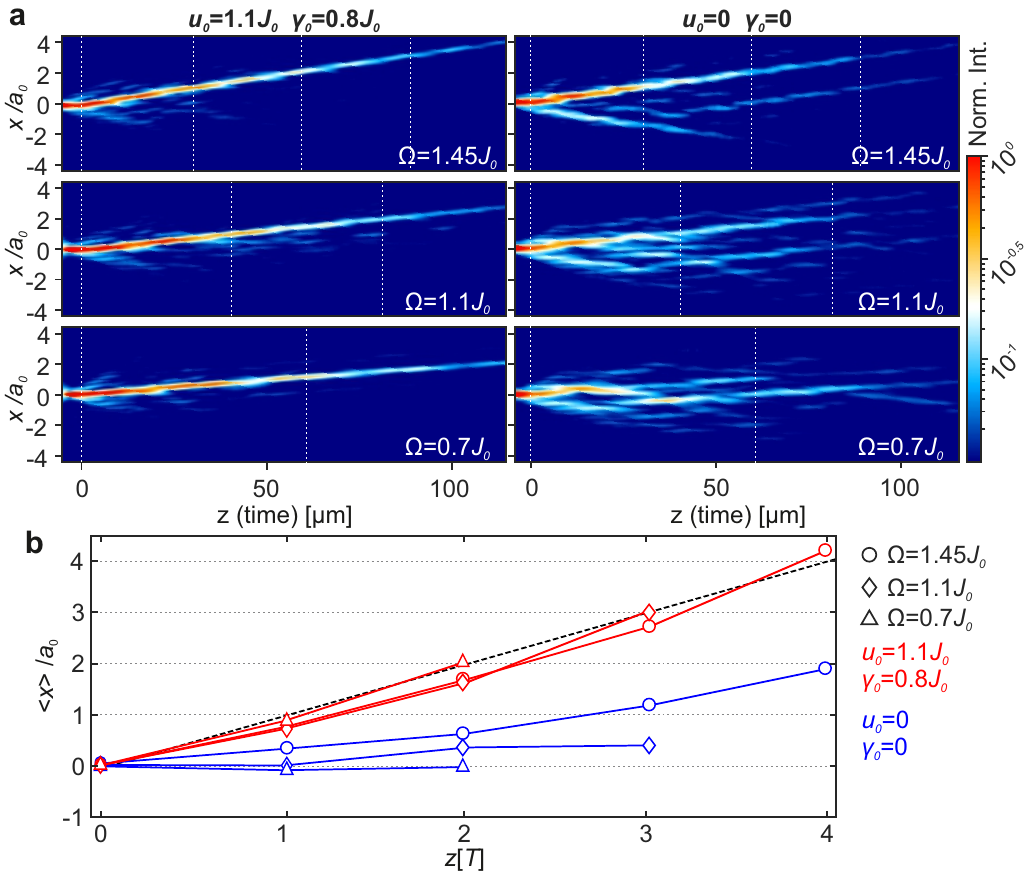}
	\caption{{\textbf{Influence of driving frequency on the transport.} (a) Real-space SPP intensity distributions for different driving frequencies and single-site excitation at waveguide $A$.  The left and right column correspond to arrays with cross-section modulation ($u_0=1.1J_0, \gamma_0=0.8$) and without cross-section modulation ($u_0=0$, $\gamma_0=0$), respectively. (b) The CoM position of the SPP intensity in dependence on propagation distance $z$ calculated from the experimental results shown in (a). Note that the $z$-axis here is normalized to the period $T$.  Red markers correspond to arrays with cross-section modulation and blue markers correspond to no modulation. The black dashed line shows the anticipated adiabatic behavior.}}
	\label{fig:ComparisonToDirectionalCouplers}
\end{figure} 

\begin{figure}[t]
\includegraphics[width=\linewidth]{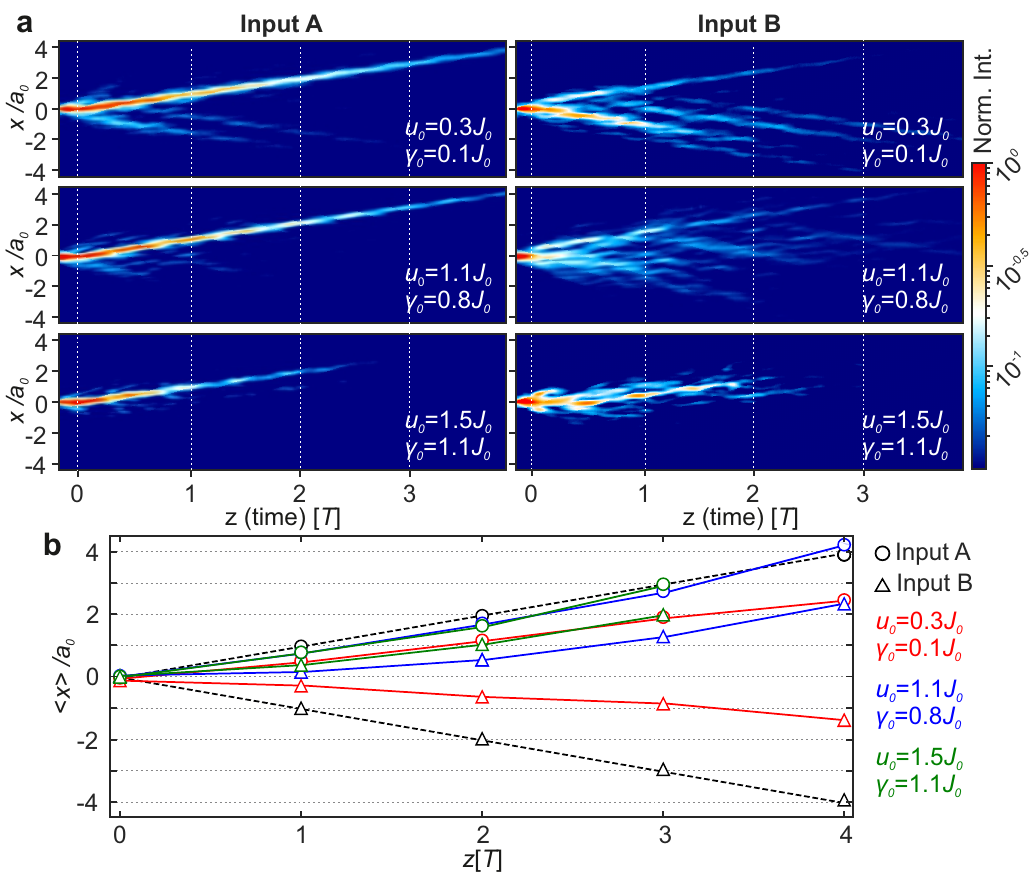}
 \caption{{\textbf{Influence of cross-section modulation and input conditions on the transport.} (a) Real-space SPP intensity distributions  for the arrays with different strengths of the cross-section modulation ($u_0=0.3J_0$, $\gamma_0=0.1 J_0$), ($u_0=1.1J_0$, $\gamma_0=0.8J_0$), and ($u_0=1.5J_0,\gamma_0=1.1J_0$). Measurements on the left-hand side show the SPP propagation after excitation at sublattice A (low-loss input). Measurements on the right-hand side show the SPP propagation after excitation at sublattice B (high-loss input) (b) The CoM position of the SPP intensity in dependence on propagation distance $z$ calculated from the experimental results shown in (a). Note that the $z$-axis here is normalized to the period $T$.  The black dashed line shows the anticipated adiabatic behavior.}}
 \label{fig:DifferentInputs}
\end{figure}  

Directional transport of light in periodically curved waveguides can be in principle also achieved by using a simple combination of directional couplers with constant effective mode index, i.e., constant waveguide cross-section~\cite{longhi2009rectification,dreisow2013spatial}. However, due to periodic exchange of power between two coupled waveguides this effect has a resonant character and the period of modulation plays in this case a crucial role.
In order to demonstrate that the directional transport in our system has a different origin, we repeat the experiment shown in Fig.\,\ref{fig:Experiment_LRM} (a) for three different driving frequencies $\Omega$ ($0.7J_0, 1.1 J_0, 1.45J_0$). Moreover we prepare two sets of samples, one with modulation of the waveguide cross-section as before ($u_0=1.1J_0$, $\gamma_0=0.8 J_0$) and the second with constant cross-section ($u_0=0$, $\gamma_0=0$). 
The measured real space intensity distributions are depicted in Fig.\,\ref{fig:ComparisonToDirectionalCouplers}~(a).
We extract from this data the CoM position after up to 4 complete periods as displayed in Fig.~\ref{fig:ComparisonToDirectionalCouplers}~(b). In the case with cross-section modulation (red markers) the CoM is shifted by one unit cell per period $T$ at all chosen driving frequencies. 
We note that the somewhat lower than unit slope of the CoM plots in Figs. 6 (b), 7 (b)  during the first pumping cycle is an artifact which arises from non-ideal excitation conditions, such as weak excitation of the neighboring waveguides. The deviations at large distance are statistical and result from increasing measurement errors due to camera noise and decaying signal intensity.
In Fourier space changing the modulation frequency influences primarily the width of the Floquet BZ: the lower is the frequency, the smaller is the distance between the neighboring bands and the smaller is the tilt of these bands which reflects the wavepacket group velocity in absolute values  (see Supplementary Figure 1).

Without cross-section modulation (blue markers in Fig.~\ref{fig:ComparisonToDirectionalCouplers}~(b)) the CoM displacement per period at these frequencies is much smaller than in the quantized case and depends on the driving frequency. These measurements confirm that the observed directional transport in our system is not a resonant directional coupler effect.

Up to now we only considered experiments with excitation at sub-lattice A (low loss input). The numerical calculations predict that the transport in the opposite direction for single site excitation at sub-lattice B is strongly suppressed by the time-periodic losses. 
To test this, we perform additional experiments to study how the transport properties depend on the initial conditions for different strengths of cross-section modulation.
In doing so we tune the amplitude of the on-site potential $u_0$ and simultaneously the loss amplitude $\gamma_0$. 
Fig.~\ref{fig:DifferentInputs}~(a)
shows the real space intensity distributions
for the excitation at the waveguides A (left column) and B (right column) for three different cross-section modulations and the driving frequency $\Omega=1.45 J_0$. 
The CoM displacement derived from this data is depicted in
Fig.~\ref{fig:DifferentInputs}~(b) (waveguide A: circles, waveguide B: triangles). In case of  small modulation strength ($u_0=0.3J_0$, $\gamma_0=0.1 J_0$, red markers) SPPs excited at site A and B are transported in +$x$ and -$x$ directions, respectively. However, for both inputs the mean displacement of the CoM is less than 1 unit cell per period. For the modulation strength ($u_0=1.1J_0$, $\gamma_0=0.8J_0$, blue markers) input A shows quantized displacement of the CoM while the sign of the mean displacement for input B switches from + to -. This effect becomes even stronger at higher modulation strength $u_0=1.5J_0,\gamma_0=1.1J_0$ (green) -- as predicted by theory (compare with Fig.\,\ref{fig:Centerofmass}~(a-b)). In Fourier space increasing the modulation strength results in a strong band broadening caused by a growing damping rate. This effect  is more pronounced for the input B (see Supplementary Figure 2).
\\
\section*{Discussion}
In this work, we introduced the concept of time-periodic dissipation in
Floquet topological systems. The theoretical analysis required the
generalization of Floquet theory to quantum mechanics with non-Hermitian,
time periodic Hamiltonians. Such quantum systems can be simulated experimentally
in dielectric-loaded surface-plasmon polariton waveguide (DLSPPW) arrays.
Specifically, we considered a non-Hermitian extension of the periodically
driven Rice-Mele model. While fast driving of dissipationless
systems always destructs the quantization of Thouless pumping, 
we predicted theoretically that time- and space-periodic dissipation can
lead to the restoration of quantized transport for nonadiabatic driving
conditions. 
This finding results from the fact that periodic loss can modify the
Floquet-Bloch band structure in such a way that the band gaps present in the
non-lossy Floquet-driven system close. In this way, a chiral Floquet band
is established that winds around the two-dimensional Floquet-Bloch Brillouin zone, 
and which thus carries quantized transport given by the Chern number. 
We emphasize that this is not merely due to a dissipation-induced band
smearing, but a true renormalization of the real part of the energy
eigenvalues, induced by the nonlinearity of the eigenvalue equation.
In a real-space picture, the phenomenon of gap closing can be understood
as selective suppression of one of the counter-propagating states.
In order to examine the theoretical predictions, we used evanescently coupled
plasmonic waveguide arrays to implement the model.
Combining real- and Fourier-space imaging, we demonstrated fast, quantized transport
in the waveguide arrays. In real space, the center of mass of the excited
surface-plasmon polariton wave packet was shifted by one unit cell per driving cycle.
In Fourier space quantized pumping is seen as a chiral Floquet band that winds around the quasienergy Brillouin zone. Additional experiments showed that, first, unlike in a simple combination of directional couplers, the SPP transport in our system is independent on the driving frequency. Second, the transport in the opposite direction is strongly suppressed. Our experimental results agree well with the theoretical predictions based on Floquet theory.

Our findings may open a new line of research using dissipative Floquet
engineering to control periodically driven quantum systems. Specifically,
it will be interesting to see whether in a conserving system time-periodic imaginary 
parts in an effective single-particle equation of motion can be induced not only by
losses but rather by interactions, and if they can be controlled so as to
establish topologically nontrivial, effective band structures. The present plasmonic waveguide setup constitutes a model for man-body systems with particle loss to an external bath. The latter are often described by the Lindblad formalism, if the bath is Markovian. The topological structure of a periodically driven system, however, becomes visible in Floquet space only. In a many-body description, this would call for a combination of the Floquet and the Lindblad techniques, which is a combination is a fundamental, unresolved problem~\cite{schnell2020there}.  Note added in proof: After submission of the final manuscript, a study on a very similar subject appeared\cite{Fehske2020}.

%-------------------------------------------------
%-------------------------------------------------
\section*{Methods}
\subsection*{Non-Hermitian Floquet theory}
In momentum space the Hamiltonian of the driven Rice-Mele model with periodic dissipation reads,
\begin{eqnarray}
	&\hat H_k(t)=(J_1\hspace*{-0.6mm}+\hspace*{-0.6mm}J_2)\cos\frac{ka_0}{2}\,\sigma_x+
	(J_1\hspace*{-0.6mm}-\hspace*{-0.6mm}J_2)\,\sin\frac{ka_0}{2}\,\sigma_y
	\nonumber\\
	&+(u_a-\mathrm{i}\gamma_a)
	(1\hspace*{-0.6mm}+\hspace*{-0.6mm}\sigma_z)/2+(u_b-\mathrm{i}\gamma_b)(1\hspace*{-0.6mm}-\hspace*{-0.6mm}\sigma_z)/2, \label{eq:MomentumHamiltonian} 
\end{eqnarray}
where the coefficients have the above time dependence, $\sigma_x$,
$\sigma_y$, $\sigma_z$ are the Pauli matrices acting in $(A,B)$ sublattice space, and $k$ and $a_0$ denote momentum and the lattice constant, 
respectively.

We now develop the Floquet formalism for non-Hermitian, periodic Hamiltonians.
Due to time periodicity, the eigenstates of $\hat H_k$ obey the Floquet
theorem \cite{Eckardt2017,Chen2018,gomez2013floquet},
\begin{equation}\label{eq:FloquetTheorem}
	|\Psi_{k\alpha}(t)\rangle=\mathrm{e}^{-\mathrm{i}\varepsilon_{k\alpha}t}|\phi_{k\alpha}(t)\rangle,
\end{equation}
where a Greek index $\alpha\in \{1,\,2\}$ denotes the band quantum 
number originating from the two sublattices, and 
$|\phi_{k\alpha}(t)\rangle = |\phi_{k\alpha}(t+T)\rangle$ are time-periodic
states which, by construction, obey the Floquet equation 
\begin{equation}\label{eq:FloquetEquation} 
	\mathcal{H}_k(t)|\phi_{k\alpha}(t)\rangle =\varepsilon_{k\alpha}|\phi_{k\alpha}(t)\rangle,   
\end{equation}
with $\hat{\mathcal{H}}_k(t):=[\hat{H}_k(t)-\mathrm{i}\partial_t]$.
The non-Hermiticity is accounted for by complex 
quasienergies $\varepsilon_{k\alpha}$. 
Note that for nonlinear or interacting, dissipative systems the
Floquet theorem would generally not hold due to non-periodic,
decaying density terms in the Hamiltonian.
Expanding the $|\phi_{k\alpha}(t)\rangle$ in the basis of time-periodic 
functions, $|\phi_{k\alpha}(t)\rangle=
\sum_{n}\mathrm{e}^{-\mathrm{i}n\Omega t}|u_{k,\alpha}^{\phantom{k,}n}\rangle$ (Floquet representation), 
Equation~(\ref{eq:FloquetEquation}) takes the form of a discrete matrix  
Floquet-Schr\"odinger equation, 
\begin{equation}\label{eq:FloquetComponents}
	\begin{aligned}
		\sum_{l,\gamma}
		(\mathcal{H}_k)^{nl}_{\beta\gamma}\,u_{k,\gamma\alpha}^{\phantom{k,}l m} = \varepsilon_{k\alpha}\,u_{k,\beta\alpha}^{\phantom{k,}nm}\,,
	\end{aligned}
\end{equation}
where $(\mathcal{H}_k)^{nl}_{\beta\gamma} =
[(H_k)^{nl}_{\beta\gamma} - n\Omega\, \delta^{nl}\delta_{\beta\gamma}]$
is the time-independent Floquet Hamiltonian and $(H_k)^{nl}_{\beta\gamma}$ the
representation of $\hat H_{k}(t)$ in the basis of time-periodic functions,
$\{\mathrm{e}^{-\mathrm{i}n\Omega t}| n\in \mathbb{Z} \}$. Equation~(\ref{eq:FloquetComponents})  determines the eigenvalues $\varepsilon_{k\alpha}$ and 
the eigenvector components
$u_{k,\beta\alpha}^{\phantom{k,}nm}\in\mathbb{C}$ 
for the above Floquet expansion. Since there are as many eigenvectors 
as the dimension of the Floquet Hamiltonian, these components not only 
carry a RM sublattice index $\beta$ and a Floquet expansion index $n$, but 
also a band index $\alpha$ and a Floquet index $m$ to label the different 
eigenvectors. Thus, $(u_{k,\beta\alpha}^{\phantom{k,}nm})$ is the matrix
comprised of column eigenvectors of Equation~(\ref{eq:FloquetComponents}). 
Note that Equation~(\ref{eq:FloquetComponents}) cannot be diagonalized
separately in the sublattice space ($\beta\alpha$) and in the
Floquet space ($nm$) because of the entanglement of both spaces. 

In quantum mechanics, expectation values of an observable $\hat{A}$ are 
calculated as overlap matrix elements like $\bra{\Psi}\hat{A}\ket{\Psi}$,
which defines the standard scalar product in Hilbert space. 
However, the eigenstates of a NH Hamiltonian $\hat{H}_k$ are generally not 
simultaneously eigenstates of $\hat{H}_k^{\dagger}$ 
\cite{brody2013biorthogonal,ibanez2011shortcuts}. As a consequence, they 
do not constitute an orthonormal basis with respect to the
standard scalar product of quantum mechanics. This hampers the expansion 
of a quantum state $\ket{\Psi(t)}$, prepared with a given initial 
condition $\ket{\Psi(t=0)}$ as in the experiments, in terms of 
Hamiltonian eigenstates.
For the sake of orthonormal basis expansions, a scalar product in 
Hilbert space can be defined by constructing the dual (bra)
states $\bra{\widetilde{\Psi}_{k\alpha}}$
corresponding to the ket states $\ket{\Psi_{k\alpha}}$ in the following way.
When $\ket{u_{k\alpha}^{m}}$ is an eigenstate of  
Equation~(\ref{eq:FloquetComponents}), it is clear that there exists an, 
in general different, adjoint state $\ket{\widetilde u_{k,\alpha}^{\phantom{k,}m}}$ 
such that 
\begin{equation}\label{eq:FloquetPartner} 
	\mathcal{H}_k^{\dagger}\ket{\widetilde u_{k,\alpha}^{\phantom{k}m}}
	=\varepsilon_{k\alpha}^*\ket{\widetilde u_{k,\alpha}^{\phantom{k}m}}.   
\end{equation}
The dual state is then obtained as $\bra{\widetilde u_{k,\alpha}^{\phantom{k}m}}=
\ket{\widetilde u_{k,\alpha}^{\phantom{k}m}}^{\dagger}$, defining the scalar 
product as $\braket{\widetilde u_{k,\alpha}^{\phantom{k}m}| u_{k',\beta}^{\phantom{k',}n}}$.
Using Equation~(\ref{eq:FloquetComponents}) and the Hermitian conjugate of
Equation~(\ref{eq:FloquetPartner}), it is easy to show that the Floquet states 
fulfill the biorthonormality (and corresponding completeness) relation 
(for non-degenerate $\varepsilon_{k\alpha}\neq\varepsilon_{k'\beta}$)
\begin{equation}\label{eq:biorthonormality}
	\braket{\widetilde u_{k\alpha}^{\phantom{k}m}|u_{k'\beta}^{\phantom{k,}n}}=
	\sum_{l,\gamma} (\widetilde u_{k,\alpha\gamma}^{\phantom{k,}ml})^*\,
	u_{k',\gamma\beta}^{\phantom{k',}l n} =\delta_{kk'}\delta_{\alpha\beta}\delta^{mn}\,.
\end{equation} 
The retarded Green's function to the NH Hamiltonian $\mathcal{H}_k$
is then the causal part of the time evolution
operator in Floquet representation,
\begin{equation}\label{eq:Greensfct}
	G_{k,\beta\alpha}^{nm}(t-t')=-\mathrm{i}\Theta(t-t')\bra{\widetilde u_{k\beta}^n}e^{-\mathrm{i}\mathcal{H}_k(t-t')}\ket{u_{k\alpha}^m},
\end{equation} 
which yields the spectral representation 
\begin{equation}\label{eq:Spectral}
	G_{k,\beta\alpha}^{\phantom{k,}nm}(E)=\sum_{l,\gamma} \frac{(\widetilde u_{k,\beta\gamma}^{\phantom{k,}nl})^*\,u_{k,\gamma\alpha}^{\phantom{k,}lm}}{E-\varepsilon_{\gamma}-l\Omega+\mathrm{i}0}\,.
\end{equation} 
Note that the lossy dynamics ($\mathrm{Im}\,\varepsilon_{k\alpha}\leq 0$)
ensures the convergence of the Fourier integral.

An arbitrary state $\ket{\Psi(t)}$ can now be expanded in the 
basis of Floquet states as 
\begin{equation}\label{eq:expansion}
	\ket{\Psi (t)}\hspace*{-0.1cm} =\hspace*{-0.1cm} \sum_{k,\alpha, n} {C}_{k\alpha}^{n} \mathrm{e} ^{-\mathrm{i}(\varepsilon_{k\alpha}+n\Omega)t} 
	\ket{u_{k,\alpha}^{\phantom{k,}n}},\ \
	{C}_{k\alpha}^{n}\hspace*{-0.1cm}=\hspace*{-0.1cm}\braket{\widetilde u_{k,\alpha}^{\phantom{k,}n}|\Psi(0)},
\end{equation}
where the time-independent expansion coefficients ${C}_{k\alpha}^{n}$ 
are calculated at the initial time $t=0$ using the biorthonormality relation
(\ref{eq:biorthonormality}) and, thus, incorportate the initial conditions on $\ket{\Psi(t)}$. 

Using the expansion (\ref{eq:expansion}), physical expectation
values for time-evolving states can now be calculated
in a straight-forward way and decay exponentially in time due to the lossy dynamics of the system. For instance, the density of a driven-dissipative Floquet state reads,
\begin{equation}\label{eq:densitydecay} 
\braket{\Psi_{k\alpha}(t)|\Psi_{k\alpha}(t)}= {\rm e}^{-\Gamma_{k\alpha}t},
\end{equation}
with the decay rate $\Gamma_{k\alpha}=-2\mathrm{Im}\varepsilon_{k\alpha}>0$.
In our DLSPPW experiments below it is
possible to directly measure the momentum- and energy-resolved population
density, i.e., intensity of the Fourier transform 
$\ket{\Psi_{k}(E)}$, which reads,
\begin{eqnarray} \label{eq:Intensity}
	&&I(E,k)= 
	\braket{\Psi_{k}(E)|\Psi_{k}(E)} \\
	&&= \sum_{n,m,\alpha\beta}\sum_{l,\gamma}  
	\frac{ C_{k\beta}^{l\ *} C_{k\alpha}^{l}\ (u_{k\beta\gamma}^{\phantom{k,}nl})^* u_{k,\gamma\alpha}^{\phantom{k,}lm}}{(E-\varepsilon_{k\beta}^*-l\Omega -\mathrm{i}0)(E-\varepsilon_{k\alpha}-l\Omega +\mathrm{i}0)}. \nonumber
\end{eqnarray}
It is seen that, in general, this expression involves the mixing of
the RM bands ($\alpha,\ \beta$), leading to
a broad spectral distribution in the FBBZ.
A distribution of this type is shown in
Fig.~\ref{fig:Quasienergies} (c). It is also possible to effectively
populate only one RM band $\alpha$ by
populating at the initial time $T=0$ one single site of the initially
nonlossy sublattice, see Fig.~\ref{fig:Quasienergies} (a) and (b).
In this case, the $\alpha-\beta$ cross terms vanish, and
Equation~(\ref{eq:Intensity}) simplifies to 
\begin{eqnarray}
	I_{\alpha}(E,k)&=&
	\braket{\Psi_{k\alpha}^{\phantom{k,}n}(E)|\Psi_{k\alpha}^{\phantom{k,}n}(E)} \nonumber\\
	&=& \sum_{n,m,l,\gamma} C_{k\alpha}^{l\ *} C_{k\alpha}^{l}  
	\frac{(u_{k\alpha\gamma}^{\phantom{k,}nl})^* u_{k,\gamma\alpha}^{\phantom{k,}lm}}{|E-\varepsilon_{k\alpha}-l\Omega|^2}. 
	\label{eq:Intensity1}
\end{eqnarray}
This is similar, albeit not identical, to the spectral density 
obtained from the imaginary part of the Green's function in
Equation~(\ref{eq:Greensfct}). Thus, measurements of the population density $I(E,k)$
of a wave function initialized at $t=0$ provide detailed information about the
stationary spectral function.

\subsection*{Dissipative transport quantization}\label{sec:TransportQuantization}

For an adiabatic Thouless pump the number of particles 
transported by one lattice constant per cycle is given by the Berry phase, 
i.e., the Berry flux penetrating a closed loop in Hamiltonian 
parameter space. Therefore it is quantized and time plays no role
\cite{Thouless1983}. 
 For fast driving, considering the states localized on single sites as approximate eigenstates for small hopping amplitude, the driving-induced hopping to neighboring sites can be viewed as Landau-Zener tunneling in real space.\cite{torosov2013non} However, the topological nature of the process is better analyzed by working in momentum and frequency space. Namely, any nonzero driving frequency $\Omega$ turns the problem into an effectively
two-dimensional (2D) one due to the periodicity in space and time. 
In this case, the Hermitian RM model possesses two counterpropagating 
chiral Floquet bands in the 2D FBBZ  
$\{-\Omega/2\leq\varepsilon< \Omega/2;-\uppi/a_0\leq k<
\uppi/a_0\}$~\cite{Kitagawa2010,titum2016anomalous}, as depicted 
in Fig.~\ref{fig:Quasienergies} (a). Quantized tranport in a Floquet band
is controlled by the winding or Chern number of the band around the FBBZ
\cite{privitera2018nonadiabatic}.
Here we investigate transport quantization in a general, fast pumped,
dissipative situation. The velocity operator reads
$\hat{v}={\rm Re}\,d\hat{\mathcal{H}_k}/dk$  ($\hbar=1$)\cite{privitera2018nonadiabatic},
i.e., for each $k$-state $\ket{\Psi_{k\alpha}(t)}$, its eigenvalue 
is the group velocity $d{\rm Re}\,\varepsilon_{k\alpha}/dk$. 
Thus, the spatial displacement
of the particle number during one pumping cycle carried by a single 
Floquet state $\ket{\Psi_{k\alpha}(t)}$ 
with a loss rate $\Gamma_{k\alpha}$ is given by
\begin{equation}\label{eq:displacement_k}
\int_{0}^{T}dt \frac{\bra{\Psi_{k\alpha}(t)}\hat{v}\ket{\Psi_{k\alpha}(t)}}{\braket{\Psi_{k\alpha}(t)|\Psi_{k\alpha}(t)}}=
\frac{d{\rm Re}\,\varepsilon_{k\alpha}}{dk}T .
\end{equation}
Note that the velocity expectation value is normalized by the exponentially 
decaying probability density, Equation~(\ref{eq:densitydecay}), such
that in Equation~(\ref{eq:displacement_k}) the exponential decay factor 
$\exp(-\Gamma_{k\alpha}t)$ drops out.
The shift per cycle carried by a band $\alpha$ with population density 
$I_{\alpha}(E,k)$ (c.f. Equation~(\ref{eq:Intensity})) 
is obtained by integrating over the FBBZ, that is, over the energy $E$ and 
all $k$ states, and reads,
\begin{equation}\label{eq:disspumping1}
L_{\alpha} = \int_{\rm{FBBZ}} \frac{dE}{\Omega}\int_{-\uppi\hspace*{-0.4mm}/\hspace*{-0.4mm}a_0}^{\uppi\hspace*{-0.4mm}/\hspace*{-0.4mm}a_0}\frac{dk}{2\uppi/a_0}\, I_{\alpha}(E,k)\,
\frac{d{\rm Re}\,\varepsilon_{k\alpha}}{dk}T.
\end{equation}
For a homogenously filled band, 
$\int dE\, I_{\alpha}(E,k)/\Omega=1$, this reduces to 
\begin{equation}\label{eq:disspumping2}
\frac{L_{\alpha}}{a_0} = \int_{-\uppi\hspace*{-0.4mm}/\hspace*{-0.4mm}a_0}^{\uppi\hspace*{-0.4mm}/\hspace*{-0.4mm}a_0}\frac{dk}{2\uppi} \,
\frac{d{\rm Re}\,\varepsilon_{k\alpha}}{dk}T = Z\,.
\end{equation}
For a periodically driven system, the dispersion $\rm Re\,\varepsilon_{k\alpha}$ is not only a periodic function of $k$, but its values are also periodic with period $\Omega$. That is, $\varepsilon_{k\alpha}$ is a mapping from the 1D circle onto the 1D circle,
${\rm Re}\,\varepsilon_{\alpha}: S^1 \to S^1$, and wraps around the 2D torus of the FBBZ as shown in Fig.~\ref{fig:Quasienergies} (d). Equation~(\ref{eq:disspumping2}) is the definition of the winding number around 
the circle. It is seen that it assumes nonzero, integer values if the dispersion continuously covers the entire FBBZ in the frequency direction,
i.e., if it is gapless, $\int dk d{\rm Re}\,\varepsilon_{k\alpha}/dk = Z\Omega=Z\,\frac{2\uppi}{T}$, since
$\varepsilon_{\uppi\hspace*{-0.4mm}/\hspace*{-0.4mm}a_0,\,\alpha}=\varepsilon_{-\uppi\hspace*{-0.4mm}/\hspace*{-0.4mm}a_0,\,\alpha}$. 
This proves the second equality in Equation~(\ref{eq:disspumping2}) for a gapless dispersion and indicates transport quantization.

\subsection*{Samples}
The DLSPPW arrays are fabricated by negative-tone gray-scale electron beam lithography (EBL)~\cite{block2014bloch,bleckmann2017spectral}. The waveguides consist of poly(methyl methacrylate) (PMMA) ridges deposited on top of a 60~nm thick gold film evaporated on a glass substrate. 
The mean center-to center distance between the ridges is $1.7~\mathrm{\upmu m}$ and the maximum deflection from the center is $0.5~\mathrm{\upmu m}$. The resulting variation of coupling constants is $J_1(z)=J_0\mathrm{e}^{-\lambda(1-\sin{\Omega z })}$, $J_2(z)=J_1(z-T/2)$ with $J_0=0.144~\mathrm{\upmu m}^{-1}$ and $\lambda=1.75$. 

The cross-section of each waveguide is controlled by the applied electron dose during the lithographic process. By varying the electron dose along the z-axis we modulate the waveguides' cross-sections and hence the propagation constants as $\beta_{a}(z)\approx\bar{\beta}- u_0\cos{(\Omega z+ \varphi)}-\mathrm{i}\gamma_a(z)$, $\beta_{b}(z)=\beta_{a}(z-T/2)$, where $\bar{\beta}=6.62+\mathrm{i}0.015~\mathrm{\upmu m}^{-1}$ corresponds to the mean height $100~\mathrm{nm}$ and the mean width $250~\mathrm{nm}$ of a waveguide and $\gamma_a(t)\approx-\gamma_0 \Theta(u_a(z))\cos(\Omega z+\varphi)$ is the periodic loss rate induced by coupling to free SPPs.
The choice of such geometrical parameters is motivated by the fact that strong losses due to coupling to continuum of free propagating SPPs occur when the height and the width of a waveguide are smaller than the corresponding mean values, i.e., $\beta_j(z)<\bar{\beta}$. Other sources of losses  can be assumed to be independent of $z$ because their variation is negligibly small in comparison to this effect.

\subsection*{Leakage radiation microscopy}
SPPs are excited by focusing a TM-polarized laser beam with free space wavelength $\lambda_0$=980~nm  (NA of the focusing objective is 0.4) onto the grating coupler deposited on top of the central waveguide (either sublattice $A$ or $B$). The propagation of SPPs in an array is monitored by real- and Fourier-space leakage radiation microscopy ~\cite{drezet2008leakage,cherpakova2017transverse}.
For this purpose, we use an oil immersion objective (63× magnification, NA=1.4)  to collect the leakage radiation. Real space intensity distributions are recorded by imaging the sample plane  onto a CMOS camera. The corresponding Fourier images are obtained by imaging the back-focal plane of the objective onto the camera. The directly transmitted laser beam is blocked by Fourier filtering.
We note that we work in the single-mode waveguide regime for all cross sections used in the experiments at the design wavelength. 

\section*{Data Availability} 
The data that support the findings of this study are available from the corresponding
author upon reasonable request. All these data are directly shown in the corresponding figures without further processing.

\bibliographystyle{apsrev4-1}
\bibliography{bibliography}

\section*{Acknowledgments}
This work was supported by the Deutsche Forschungsgemeinschaft (DFG)  within  SFB/TR  185  (277625399) and the Cluster of Excellence ML4Q (390534769). 

\section*{Author contributions}
ZF fabricated the samples, conducted the experiments, and performed the numerical calculations with the help of HQ. SL and JK conceived the project and supervised ZF and HQ, respectively.
All authors contributed to the discussion and interpretation of the results as well as writing the manuscript. 

\section*{Competing interests}
The authors declare no competing financial interests.

\end{document}